# One-dimensional Dynamics for a Discontinuous Singular Map and the Routes to Chaos

Moorad Alexanian

*Department of Physics and Physical Oceanography*
*University of North Carolina Wilmington, Wilmington, NC 28403-5606*

Email: alexanian@uncw.edu

**Abstract**. We visit a previously proposed discontinuous, two-parameter generalization of the continuous, one-parameter logistic map and present exhaustive numerical studies of the behavior for different values of the two parameters and initial points $x_0$. In particular, routes to chaos exist that do not exhibit period-doubling whereas period-doubling is the sole route to chaos in the logistic map. Aperiodic maps are found that lead to cobwebs with $x = \pm\infty$ as accumulation points, where every neighborhood contains infinitely many points generated by the map.

## 1. Introduction

The one-dimensional logistic map $x_{n+1} = \lambda x_n(1-x_n)$ illustrates the period-doubling route to chaos [1, 2]. A detailed history of iterated maps is given by Wolfram [3]. More recently, the logistic map is used also to exhibit the effect of chaos on the description of intrinsic fluctuations by constructing a master map equation for the logistic map [4]. There is also interest in discontinuous logistic maps where inverse cascade arises [5] and where also direct cascade occurs [6]. It is interesting that the logistic map has been extended into the quantum realm by deriving a logistic map with quantum corrections by coupling a kicked quantum system to a bath of harmonic oscillators [7]. In a new proposal for a quantum key distribution, the quantum logistic map is used as a pseudo-random number generator for photon polarization state measurement bases choice [8].

In this paper we consider a discontinuous map which, however, is also singular and was introduced some time ago [9]. This paper is arranged as follows. In Sec. 2, we review the discontinuous, one-dimensional map its fixed points and their stability conditions. In Sec. 3, the approach to chaos is studied for several cases involving stable and unstable fixed points for differing values of the parameter $\alpha$. In Sec. 4, we present an aperiodic map with accumulation points at $x = \pm\infty$. Finally, Sec. 5 summarizes our results.

## 2. Discontinuous one-dimensional map

In what follows, we consider the two-parameter, one-dimensional discontinuous map [9]

$$x_{n+1} = \mu x_n \frac{(1-x_n)}{\alpha - x_n} \tag{1}$$

with $\mu$ and $\alpha$ real. In the limit $|\mu| \to \infty$ and $|\alpha| \to \infty$ such as $\mu/\alpha \to \lambda$, with $-\infty < \lambda < \infty$, map (1) reduces to the logistic map

$$x_{n+1} = \lambda x_n (1 - x_n) \tag{2}$$

The fixed points of (1) are $x^* = 0$ and $x_0^* = (\mu - \alpha)/(\mu - 1)$ with stability criteria $|\mu/\alpha| < 1$ and $|(2\mu\alpha - \mu^2 - \alpha)/\mu(\alpha - 1)| < 1$, respectively. All maps (1) with parameters $\alpha$ and $\mu$ in the wedge $0 < \mu < \alpha$ give rise to a stable fixed point at $x^* = 0$.

Under the transformation

$$x_n = \alpha/u_n, \tag{3}$$

map (1) becomes

$$u_{n+1} = \frac{\alpha}{\mu} \frac{u_n(1-u_n)}{\alpha - u_n}. \tag{4}$$

Accordingly, map (1) is invariant under the transformation (3) for $\alpha = \mu^2$ and the above transformation is quite useful in determining the values of the iterations of map (1) for values of $0 < \alpha < \mu^2$ and $\mu > 1$ from iterations of map (1) for values $\alpha > \mu^2$ and $\mu > 1$.

## 3. Approach to chaos

### a. One stable fixed point

It was found [9] that the curve $\alpha(\mu)$ with $\mu > 1$

$$g_2(\mu) = \mu(\mu+1) / (3\mu-1) \tag{5}$$

represents the transition where the fixed $x_0^*$ is unstable and a new periodic orbit of period-2 begins. Note that in the limit $\mu \to \infty$ and $\alpha \to \infty$, one obtains $\mu/\alpha \to 3$, which corresponds to $\lambda = 3$ for the logistic map (2). This period-2 orbit becomes unstable in turn as $\alpha$ decreases further and a period-4 orbit appears when [9]

$$g_4(\mu) = \frac{-\mu(\mu^2 - \mu - 1) + \mu\sqrt{\mu(6\mu^3 + 10\mu^2 + 7\mu + 2)}}{5\mu^2 + 2\mu - 1}. \tag{6}$$

Note again that in the limit $\mu \to \infty$ and $\alpha \to \infty$, one obtains $\mu/\alpha \to 1+\sqrt{6}$, which corresponds to $\lambda = 1+\sqrt{6}$ for the logistic map (2). Of course, as $\alpha$ decreases further, a period-doubling sequence of period-$2^n$ with $n = 3, 4, \cdots$ emerge for $\mu > 1$.

It is interesting that one can find the limit of the above period-doubling sequence as $n \to \infty$ by considering the value of $\alpha$ for which the maximum value of map (1) is set equal to 1. Now the extrema occur at

$$x_{ex} = \alpha \pm \sqrt{\alpha^2 - \alpha} \tag{7}$$

with the lower value representing the maximum and the upper value representing the minimum. Note that real values of $x_{ex}$ occur only for $\alpha > 1$ or $\alpha < 0$. Accordingly, for the maximum value set equal to 1 at $x_{ex} = \alpha - \sqrt{\alpha^2 - \alpha}$ one has for the value of $\alpha$

$$g_\infty(\mu) = \frac{(\mu+1)^2}{4\mu} \tag{8}$$

and one obtains for map (1) what is equivalent to $\lambda = 4$ in the logistic map (2), where chaos ensues, viz., sensitive dependence on initial conditions. Note that in the limit $\mu \to \infty$ and $\alpha \to \infty$ with a finite ratio one obtains $\mu/\alpha \to 4$, which corresponds to $\lambda = 4$ for the logistic map (2). Note that (8) is invariant under the transformation $\mu \to 1/\mu$.

All routes, beginning at $\alpha < \mu^2$ with $\mu > 1$ in Fig. 1 and ending at $g_\infty(\mu)$, represent the period-doubling route to chaos precisely as in the logistic map in the range $0 < \lambda \leq 4$. For $\mu < \alpha < \mu^2$, $x^* = 0$ is a stable fixed point and it corresponds to the logistic map with $0 < \lambda < 1$. For $\mu(\mu+1)/(3\mu-1) < \alpha < \mu$, $x_0^* = (\mu-\alpha)/(\mu-1)$ is a stable fixed point and corresponds to the logistic map with $1 < \lambda < 3$. For $\alpha < \mu$, on obtains the period-doubling route to chaos, with Fig. 2 showing the period-doubling and Fig. 3 the chaotic behavior.

It is important to remark that the logistic map has always a stable fixed point for $-1 < \lambda < 3$ but never two stable fixed points. The latter is not the case for map (1), which can possess two stable fixed points, which leads to novel route to chaos.

The analytic continuation of the behavior of map (1), for $\alpha>1$ and $\mu > 1$, to values of $\mu < 1$ and $\alpha > 1$ is accomplished with the aid of the transformed map (4). The $g_\infty(\mu)$ is analytically continued to

$$u_{n+1} = \frac{1}{\mu} \frac{u_n(1-u_n)}{(\mu+1)^2/(4\mu) - u_n}, \tag{9}$$

for $\mu < 1$ and $\alpha > 1$, owing to the invariance of $(\mu+1)^2/4\mu$ under the transformation $\mu \to 1/\mu$. For given values of $\alpha$, (8) has two solutions, viz., $\mu$ and $1/\mu$, with their product being unity. However, for $g_{2n}(\mu)$ with $1 \le n < \infty$, the application of transformation (3) is somewhat different. For instance, for $g_2(\mu)$, the analytic continuation to $\mu < 1$ implies that

$$u_{n+1} = \frac{\mu+1}{3\mu-1} \frac{u_n(1-u_n)}{\mu(\mu+1)/(3\mu-1) - u_n}, \tag{10}$$

with the aid of (4). In order to compare the iterations of the two maps, we must consider a particular value of $\alpha$. For instance, for $\alpha = \mu(\mu+1)/(3\mu-1) = 5$, we have two solutions for $\mu$, viz., $\mu = 7 + 2\sqrt{11} \approx 13.633$ and $\mu = 7 - 2\sqrt{11} \approx 0.367$. The fixed points $x_0^*$ are 0.683 and 7.319, respectively, and are unstable since by the stability criteria of the derivative at the fixed point, viz., $|(2\alpha\mu - \mu^2 - \alpha)/(\mu(\alpha-1))| < 1$, one obtains instead a value of -1.00 for both fixed points, which leads to period-doubling.

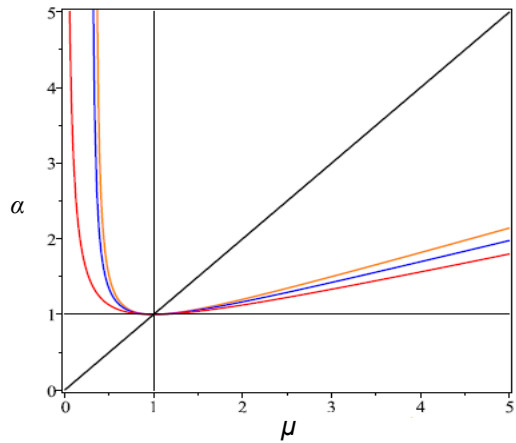

**Fig. 1.** Plots of $\alpha = g_{2n}(\mu)$, $n = 1$ (coral), $n = 2$ (blue), and $n = \infty$ (red).

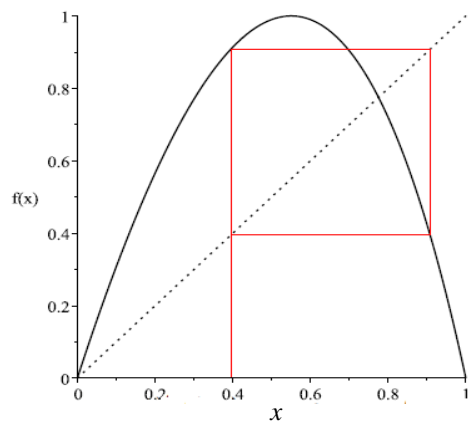

**Fig. 2.** Iteration scheme for map (1) with $\alpha = (\mu+1)^2/(4\mu) = 3$ and $\mu = 9.89897948$ leading to period-2 behavior. The two attractor points lie at the top left and bottom right of the rectangle at 0.3949806332 and 0.9080809102, respectively.

The maps $g_{2n}(\mu)$, for $\mu > 1$ and $\alpha > 1$, can be considered for values $\mu < 1$ and $\alpha > 1$. It is clear that the stability criteria indicate that $x^* = 0$ is a stable fixed point for all maps $g_{2n}(\mu)$ for $\mu < 1$ and $\alpha > 1$. At the highly singular point $\alpha = \mu = 1$ of map (1), all points $x$ is fixed points, albeit, not stable fixed points. Any route from the region with $\mu < 1$ and $\alpha > 1$ via the singular point can lead to any period-$2^n$ orbit including directly into chaos thus avoiding the period-doubling route to chaos, which is the norm in the logistic map.

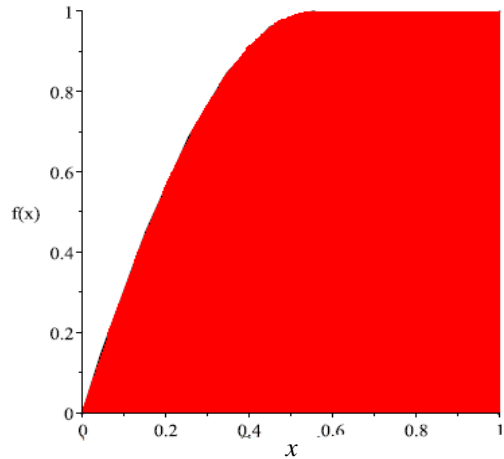

**Fig. 3.** Chaotic map (1) with $\alpha = (\mu+1)^2/(4\mu)$ with $\mu = 9.89897948$, i.e., $\alpha = 3.00$, with initial point $x_0 = 0.5$. The approach to chaos is via period-doubling. Precisely the same result is obtained with map (9) with $\mu = 0.1010205144$ with same initial point. Fig. 2 shows period-2 behavior for same parameters.

**b. Two stable fixed points: $\alpha > 0$**

An interesting feature of map (1) is the realization of maps with two stable fixed points, which cannot occur in the logistic map. Consider the map

$$x_{n+1} = \mu x_n \frac{(1-x_n)}{\frac{1}{\sqrt{\mu}} - x_n}, \qquad (11)$$

for $0 < \mu < 1$. The stable fixed points are $x^*=0$ and $x_0^*=(\mu^{3/2}-1)/[\mu^{1/2}(\mu-1)]$. The fixed point at $x^*=0$ is stable for all values $0 < \mu < 1$. The stability of the fixed point $x_0^*$ is indicated in Fig. 4, which becomes unstable for $\mu < 0.5304$. Therefore, as $\mu$ decreases we observe a period-doubling route to chaos as shown in Fig. 5. For $\mu > 1$, both fixed points become unstable. Fig. 6 shows the nature of the transition to chaos, which is quite different from the logistic map since the chaotic region encompasses a larger region of the $x$-axis.

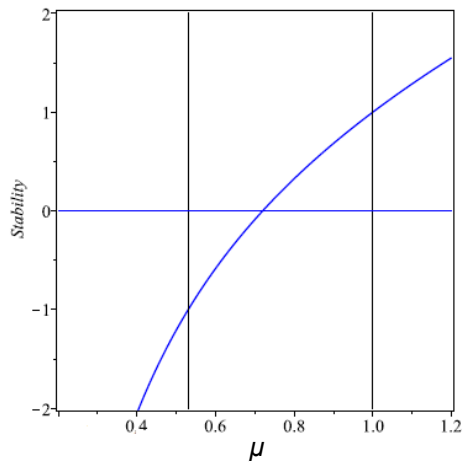

**Fig. 4.** Stability of fixed point $x_0^* = (\mu^{3/2}-1)/[\mu^{1/2}(\mu-1)]$ for map (11). Stable for $0.5304 < \mu < 1$, unstable otherwise.

**c. Two unstable fixed points: $\alpha < 0$**

One obtains interesting behavior for map (1) for negative values of α. Consider the following map

$$x_{n+1} = \frac{(3\mu-1)^2}{\mu^3(\mu+1)^2} \frac{x_n(1-x_n)}{\frac{3\mu-1}{\mu(\mu+1)} - x_n}. \qquad (12)$$

The fixed points are $x^* = 0$ and $x_0^* = (\mu^2 + 2\mu - 1)(3\mu - 1)/(\mu^4 + 3\mu^3 + 4\mu^2 - 5\mu + 1)$, the former is unstable at $\mu \approx 0.2956$ and the latter at $\mu \approx 0.3084$, both with $df(x)/dx = -1$. The iteration scheme leading to chaos is shown in Fig. 7 for initial points $x_0 > -0.2956$ for $\mu \approx 0.2956$ when $x^* = 0$ is just unstable. The iteration scheme leading to chaos is shown in Fig. 8 for $x_0 < -0.1854$ for $\mu \approx 0.3084$ when $x_0^*$ is just unstable. In both Fig. 7 and Fig. 8, the approach to chaos is via period-doubling bifurcations, where the chaotic regions are over small regions of *x*.

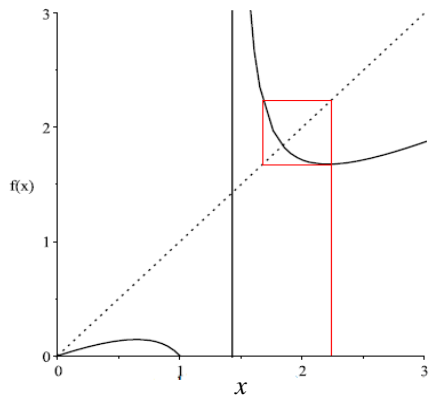

**Fig. 5.** Period-doubling for map (11) for $\mu = 0.49$ with initial point $x_0 = 2.2382$.

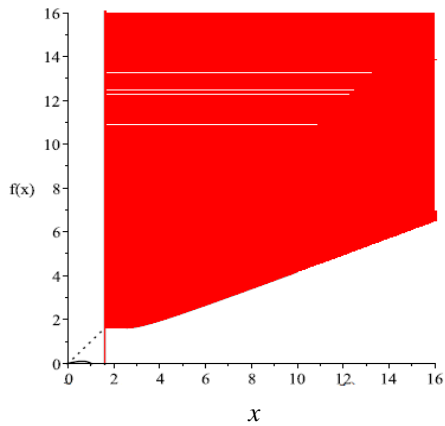

**Fig. 6.** Chaotic region for map (11) for $\mu = 0.39$.

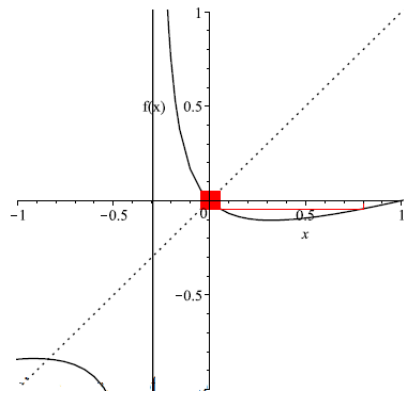

**Fig. 7.** Chaotic behavior of map (12) for $\mu \approx 0.2956$ with $x_0 = 0.8$.

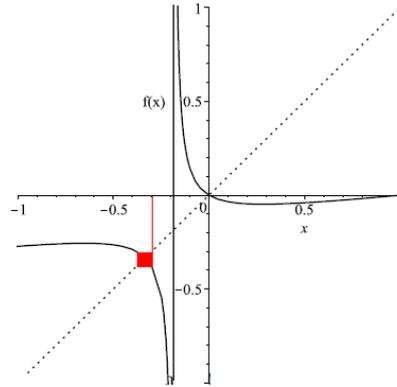

**Fig. 8.** Chaotic behavior of map (12) for $\mu \approx 0.3084$ with $x_0 = -0.3$.

### d. Two unstable fixed points: $\alpha = \mu$

Consider the following map

$$x_{n+1} = \mu \frac{x_n(1-x_n)}{\mu - x_n}, \qquad (13)$$

where $\alpha = \mu$ in (1). Both fixed points are equal to zero, i.e, $x^* = x_0^* = 0$, and $df(x)/dx = 1$. A very novel case for map (13) is obtained for $\mu \approx -0.522341$ and shown in Fig. 9. All initial point $x_0$ in the iteration process with starting points $0 \leq x_0 \leq 1$ remain in the domain $0 \leq x \leq 1$. In one-dimensional maps, one encounters fixed points, which are either attractors to which the trajectory moves to or to repellors, which repel nearby trajectories. Here we have a rather curious case where the repellor is actually all trajectories obtained from a starting point $x_0$ of the iteration that lie in the interval $0 \leq x_0 \leq 1$.

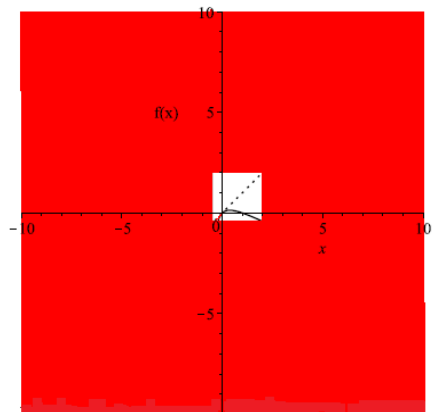

**Fig. 9.** Chaotic behavior of map (13) for $\mu \approx -0.522341$ with $x_0 = 2.0$.

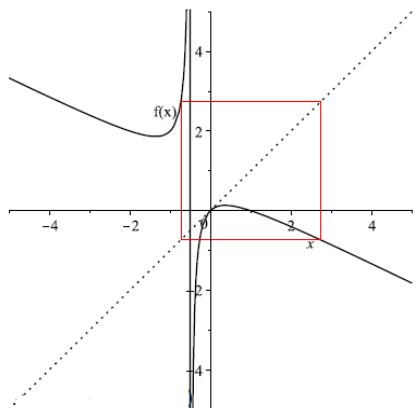

**Fig. 10**. Period-2 for map (13) for $\mu = -0.5$ and $x_0 = 2.732050808$.

Consider map (13) for $\mu=-0.5$. It is remarkable that for a fixed value of $\mu$, we can obtain a sequence of period-$n$, with $n = 2, 3, 4, 5, \cdots$, by just starting the iteration scheme at different values of the initial starting point $x_0$. Fig. 10 shows the iteration scheme leading to period-2 for $\mu=-0.5$ and $x_0=2.732050808$. Fig. 11 shows the iteration scheme leading to period-3 for $\mu=-0.5$ and $x_0=1.879385242$. Fig. 12 shows the iteration scheme leading to period-4 for $\mu=-0.5$ and $x_0=2.358929287$. Fig. 13 shows the iteration scheme leading to period-5 for $\mu=-0.5$ and $x_0=2.075558177$. Preliminary numerical iterations indicate that the chaotic state associated with a particular initial $x_0$ for $\mu=-0.5$ would be equal as that given by Fig. 9. The sequence of period-$n$ for real values of $x_0$ is true only for $|\mu|<1$ while for $|\mu|>1$ the corresponding values are imaginary numbers.

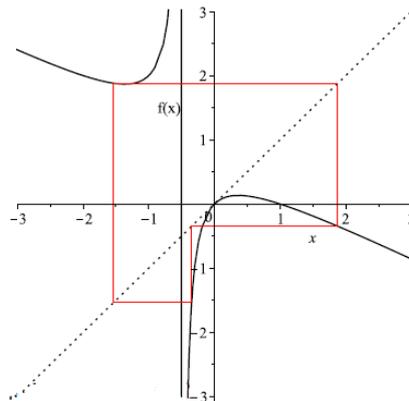

**Fig. 11.** Period-3 for map (13) for $\mu = -0.5$ and $x_0 = 1.879385242$.

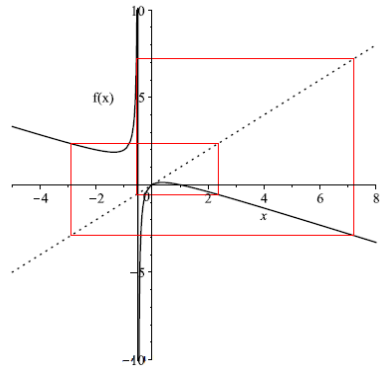

**Fig. 12.** Period-4 for map (13) for $\mu = -0.5$ and $x_0 = 2.358929287$.

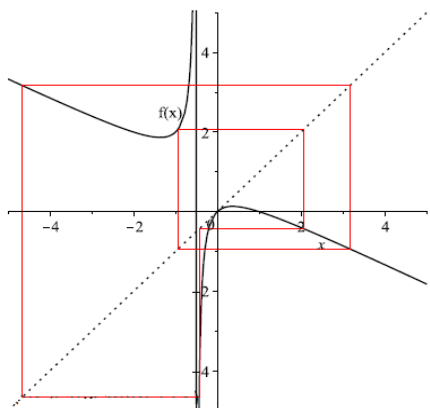

**Fig. 13.** Period-5 for map (13) for $\mu = -0.5$ and $x_0 = 2.075558177$.

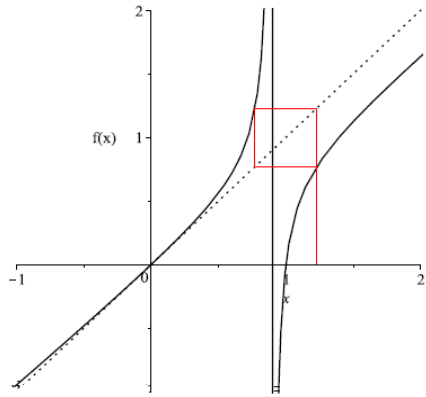

**Fig. 14.** Period-2 for map (13) for $\mu = 0.9$ and $x_0 = 1.229415734$.

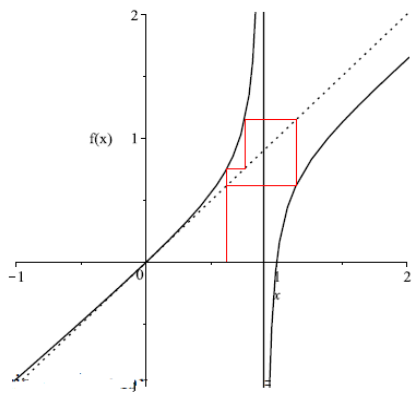

**Fig. 15.** Period-3 for map (13) for $\mu = 0.9$ and $x_0 = .6189168183$.

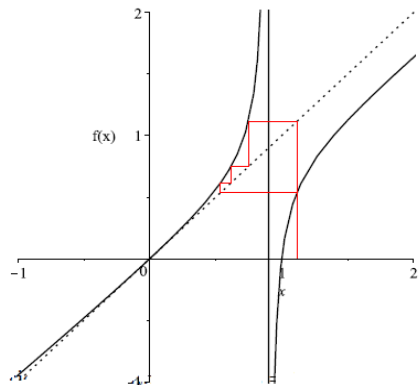

**Fig. 16.** Period-4 for map (13) for $\mu = 0.9$ and $x_0 = 1.114730637$.

Sarkovskii's theorem states that if a discrete, continuous function on the real line has a periodic point of period-3, then it must have periodic points of every other period [2]. Note that in our discontinuous, singular map (13), the presence of period-$n$, $n = 2, 3, 4, 5 \cdots$ requires only that $|\mu|<1$. Actually, map (13) is equivalent to a large class of rational functions with two unequal zeros and a simple pole.

Consider the three-parameter family of rational functions

$$x_{n+1} = \tilde{\mu} \, \frac{(x_n - a)(b - x_n)}{(c - x_n)}, \qquad (14)$$

where $a \neq b \neq c$. Map (13) corresponds to $a = 0$, $b = 1$, and $c = \tilde{\mu} = \mu$. Map (14) possesses two real fixed points for $\tilde{\mu} \neq 1$ provided

$$(a\tilde{\mu} - b\tilde{\mu} - c)^2 + 4b\tilde{\mu}(a - c) \geq 0. \qquad (15)$$

If the equality holds in (15), then there is a single fixed point at $x_0^* = \frac{\tilde{\mu}(a+b)-c}{2(\tilde{\mu}-1)}$. If $\tilde{\mu}=1$, then there is a single fixed point $\frac{ab}{a+b-c}$. at $x_0^* =$

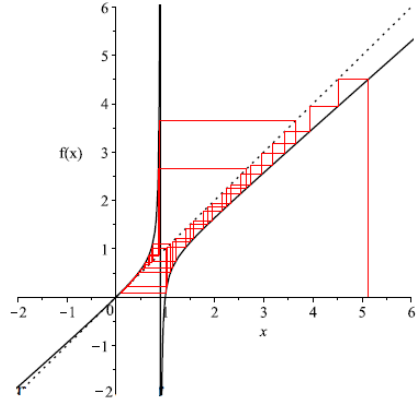

**Fig. 17.** Bouncing behavior of map (13) for $\mu = 0.9$ and $x_0 = 5.13$.

**4. Aperiodic map**

Under the transformation $x_n = a + (b - a)u_n$, (14) becomes

$$u_{n+1} = -\frac{a}{b-a} + \tilde{\mu}\frac{u_n(1 - u_n)}{(c - a)/(b - a) - u_n}. \qquad (16)$$

We conjecture that every map of the form (14) with $\tilde{\mu} = (c - a)/(b - a)$ and $|\tilde{\mu}| < 1$ has period-$n$, $n = 1, 2, 3, 4, \cdots$. However, a rather interesting and strange map results for $\tilde{\mu} = -1$, viz., $a = (b+c)/2$. Consider (14) for $a = 2$, $b = 2.5$, and $c = 1.5$ with resulting $\tilde{\mu} = (c - a)/(b - a) = -1$. One obtains from (16) that

$$u_{n+1} = -4 + u_n\frac{1 - u_n}{1 + u_n}. \qquad (17)$$

Fig. 18 shows the iterations of map (17) for the initial point $x_0 = 2.4$.

It is important to remark the difference between the maps associated with Fig. 9 and Fig. 18. The former has been verified with orbits with period-$n$, with $n = 1, 2, 4, 6, 7$, note the absence of period-3, while the latter is strictly aperiodic. This difference gives rise to quite different chaotic states. Fig 9 shows most of the real axis as the chaotic-band whereas in Fig. 18, the points $\pm\infty$ are the accumulation points of the iterations.

Our result closest to the logistic map follows from (16) when $c = 0$ and the map contains only a single parameter and so

$$u_{n+1} = \tilde{\mu} + \tilde{\mu}\frac{u_n(1 - u_n)}{\tilde{\mu} - u_n}, \qquad (18)$$

where $\tilde{\mu} = -a/(b - a)$. Map (14) contains a simple pole at $x = 0$ and so since $x_0 = a + (b - a)u_0$ one must have $x_0 \neq 0$, i.e., $u_0 \neq \tilde{\mu}$, which is the singularity in map (18). Consider map (18) for $\tilde{\mu} = 2$, which has a fixed points at $x \approx 1.2361$ and $-3.2361$ and periodic cycles with period-$n$, $n = 2, 3, 4 \cdots$. Figs. (19) and (20) show the first of these two periodic orbits.

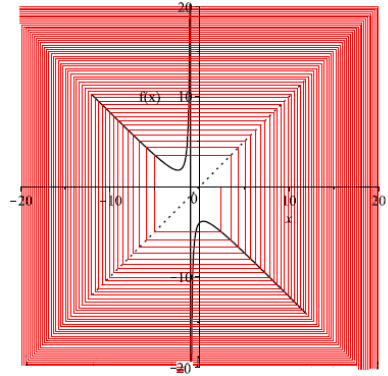

**Fig. 18.** Aperiodic behavior of map (16) for *a* = 2.0, *b* = 2.5 and *c* = 1.5 and so $\tilde{\mu} = -1.0$ with initial point $x_0 = 2.4$, viz., map (17) with same initial point.

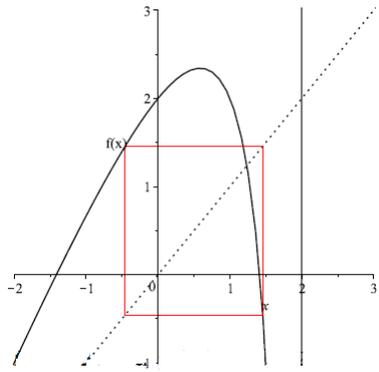

**Fig. 19.** Period-2 for map (18) for $\tilde{\mu} = 2$ and $u_0 = -0.4574$.

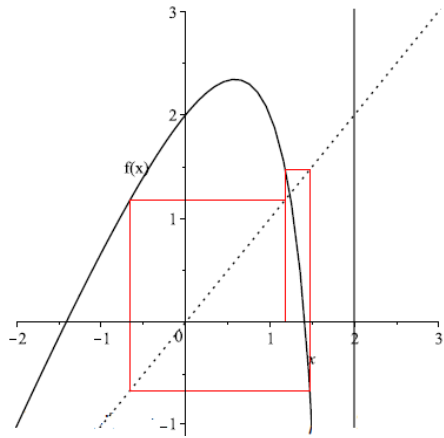

**Fig. 20.** Period-3 for map (18) for $\tilde{\mu} = 2$ and $u_0 = 1.1820$.

## 5. Conclusions

We numerically study discontinuous, singular, two-parameter maps that were analyzed only qualitatively thirty years ago. The main emphasis of this paper is to numerically study this original map and show behavior that is quite distinct from the well-known logistic map. Amongst the many difference, the more dramatic are the avoidance of the approach to chaos via period-doubling bifurcations as is the case in the logistic map. In addition, we find aperiodic maps where there are no periodic trajectory points whatsoever, instead, one has that $x = \pm\infty$ are points of accumulations of the iterations, where no matter how small the neighborhood, an unlimited number of terms of the map can be found arbitrarily close to $x = \pm\infty$.